\documentclass[preprint]{aastex63}

\usepackage{amsmath, bm}
\usepackage{appendix}
\usepackage{color}
\shorttitle{}
\shortauthors{}
\graphicspath{{./}{figure/}}
\graphicspath{{./figure/}}

\begin{document}

\title{Magnetic Flux Transport in Advection Dominated Accretion Flow Towards the Formation of Magnetically Arrested Disk}

\correspondingauthor{Jia-Wen Li \& Xinwu Cao}
\email{jwliynu@ynu.edu.cn, xwcao@zju.edu.cn}

\author{Jia-Wen Li}
\affiliation{Department of Astronomy, School of Physics and Astronomy, Key Laboratory of Astroparticle Physics of Yunnan Province, \\Yunnan University, Kunming 650091,
People’s Republic of China; jwliynu@ynu.edu.cn}

\author{Xinwu Cao}
\affiliation{ Institute for Astronomy, School of Physics, Zhejiang University,  866 Yuhangtang Road, Hangzhou 310058, \\ People’s Republic of China; xwcao@zju.edu.cn}



\begin{abstract}
The magnetically arrested disks (MADs) have attracted much attention in recent years.  The formation of MADs are usually attributed to the accumulation of a sufficient amount of dynamically significant poloidal magnetic flux. In this work, the magnetic flux transport within an advection dominated accretion flow and the formation of a MAD are investigated. The structure and dynamics of an inner MAD connected with an outer ADAF are derived by solving a set of differential equations with suitable boundary conditions. We find that an inner MAD disk is eventually formed at a region about several ten Schwarzschild radius outside the horizon. Due to the presence of strong large-scale magnetic field, the radial velocity of the accretion flow is significantly decreased. The angular velocity of the MAD region is highly subkeplerian with $\Omega \sim (0.4-0.5)\Omega_{\rm K}$ and the corresponding ratio of gas to magnetic pressure is about $\beta \lesssim 1$. Also, we find that  MAD is unlikely to be formed through the inward flux advection process when the external magnetic field strength weak enough with $\beta_{\rm out}\gtrsim 100$ around $R_{\rm out}\sim 1000R_{\rm s}$. Based on the rough estimate, we find that the jet power of a black hole, with mass $M_{\rm BH}$ and spin $a_*$, surrounded by an ADAF with inner MAD region is about two order of magnitude larger than that of a black hole surrounded by a normal ADAF. This may account for the powerful jets observed in some Fanaroff–Riley type I galaxies with a very low Eddington ratio. 
\end{abstract}


\section{introduction}\label{sec:intro}
 A magnetically arrested disks (MADs) is a strongly magnetized accretion flow where the large-scale magnetic field is dynamically important. The MADs have attracted much attention in recent years because it is suggested that MADs provide an ideal dynamic environment for powerful jet production via the extraction of the black hole spin energy\citep[][]{2003PASJ...55L..69N,2011MNRAS.418L..79T,2012MNRAS.423L..55T,2019MNRAS.487..550L}. Observational evidence from EHT (Event Horizon Telescope) suggests that M87* and Sgr A* are likely to be in a MAD state \citep[][]{2022ApJ...930L..16E,2021ApJ...910L..13E}. 
The wind-fed numerical simulations,  recently, down by \cite{2023MNRAS.521.4277R} found that Sgr A * can reach the MAD state in the inner region as the poloidal magnetic fields are efficiently transported inward from larger scales, the plasma $\beta$ can reach $\sim1-2$ at radii well outside the event horizon. Also, there are some observational evidences in favor of a MAD configuration is expected in nature around the central accreting black hole \citep[][]{2014Natur.510..126Z,2015MNRAS.449..316N,2022A&A...663L...4L,2023Sci...381..961Y,2024MNRAS.530..530H}. When enough net magnetic flux is accumulated in the inner region of the disk, the magnetic force exerted by the large-scale field can become strong enough to impede the accreting gas and lead to a MAD accretion state, in that state the accretion flow may be dispersed into magnetic confined blobs or streams, and be accreted asymmetrically with a typical radial velocity around $v_R \sim (0.001-0.01)v_{\rm ff}$ \citep[][]{2003PASJ...55L..69N}, $v_{\rm ff}$ is the free-fall velocity. The dynamics and structure of the accretion flow may be significantly altered due to presence of the large-scale magnetic field. 

The disk needs sufficient net magnetic flux to transfer into a MAD state. There are two main approaches for the net flux accumulation, i.e., flux generated in-situ and/or flux advected inward. In the picture of in situ flux generation, the large-scale poloidal magnetic flux is expected to be generated through a self-excited turbulent dynamo \citep[][]{1995ApJ...446..741B} powered by the magneto-rotational instability \citep[MRI,][]{1991ApJ...376..214B,1998RvMP...70....1B} process. The $\alpha^2$ and $\alpha-\Omega $ type dynamo are responsible for generating and amplification of the magnetic field, under the complicated magnetic reconnection effect, a large-scale ordered magnetic field is eventually generated\citep[see][for details]{2005PhR...417....1B}.  The nonlinear Local shearing box or global simulations found that the dynamo generated large-scale magnetic field is usually weak in radiatively inefficient accretion flows, and the scale of the in-situ generated field through the dynamo process is more likely limited by the scale height of the accretion flows \citep[][]{2008ApJ...678.1180B,2012MNRAS.426.3241N,2013ApJ...769...76B,2016MNRAS.460.3488S,2016MNRAS.457..857S,2018ApJ...854....6H,2020MNRAS.494.4854D}. It is still not clear whether the dynamo process operating in an astrophysical disk can generate a large-scale ordered magnetic field, which is expected to driven powerful jets/outflows and lead to an accretion flow transfer into a MAD state. More recently, however, the 3D global simulations done by \cite{2020MNRAS.494.3656L} found that the production of a large-scale poloidal field with scale height $H\propto R$ eventually led to the inner accretion flow transfer into MAD state and very power jets are observed in their simulations. But we note here that the initial condition of their simulation starts from a strong $(\beta \sim  5)$ and coherent toroidal field configuration, the formation of such an initial condition in the outer edge of the disk needs further study. For the flux-advected inward picture. \cite{1994MNRAS.267..235L} constructed a seminal model to study the magnetic flux advected in a standard thin disk, in which the turbulence, described by a $\alpha$ parameter \citep[][]{1973A&A....24..337S}, which is responsible for radial angular momentum transport. Due to the small radial accretion velocity of the thin disk, they found that it is difficult for a $\alpha$-thin disk to transport magnetic flux inward. This conclusion is limited by the approach adopted in their model calculations, i.e., they solved the height-integrated induction equation to get the global magnetic field configurations. The issue of inefficient flux transportation within a thin disk can potentially be resolved if one takes the vertical disk structure and/or the magnetic driven outflows into consideration \citep[][]{2009ApJ...701..885L,2012MNRAS.424.2097G,2013MNRAS.430..822G,2013ApJ...765..149C,2014ApJ...786....6L,2019ApJ...872..149L,2021ApJ...909..158L}. The advection dominated accretion flows \citep[ADAFs][]{1994ApJ...428L..13N,1995ApJ...452..710N} are much thicker than thin discs, hence the faster radial velocity. It is suggested that due to the rapid radial accretion velocity, ADAF can efficiently transport magnetic flux to the inner region, with the flux continuously accumulating in the vicinity of the black hole, a MAD is formed in the inner region of the ADAF \citep[][]{2011ApJ...737...94C}. Evidence for the flux advection picture is also observed in 3D global simulations \citep[][]{2009ApJ...707..428B,2011MNRAS.418L..79T,2014ApJ...784..121S,2018ApJ...857...34Z,2020MNRAS.492.1855M,2023ApJ...944..182D}. Recently, the work done by \citep[][]{2024MNRAS.530..530H} found that the mass accretion in a sample of low Eddington ratio ($\lesssim 10^{-3.4}$) Fanaroff–Riley type I (FR I) radio galaxies can be modeled as ADAF, meanwhile, the power of the detected jets systematically higher than that of the jets powered by the general Blandford–Znajek mechanism. This implies that there should be a MAD disk in the inner region of the ADAF on which the powerful FR I jets are launched. 

The physical mechanisms causing magnetic flux saturation in the MAD state and determining the steady-state structure remain for further studies. Thus, understanding how the magnetic flux is transported in the disk is important in the study of the dynamical structure and angular momentum transport in MAD. The observational evidence from  M87*, SgrA* and a series of MAD numerical simulations made it evident that the MAD regime should be reasonably common in nature, especially for the low luminosity accretion system containing an ADAF such as M87*, SgrA*, and so on. In this paper, we investigate the process of magnetic flux transport within an ADAF and the formation of a MAD disk around the central black hole. Our work is organized as follows, Section \ref{sec:model} contains the model of ADAF and the evolution of the large-scale magnetic. The numerical setup is described in Section \ref{sec:setup}. The results of the model calculations are given in Section \ref{sec:results}. The last section contains the conclusions and discussions.

\section{model}\label{sec:model}
The radial accretion velocity of an ADAF is significantly faster than that of a thin disk, which results in an efficient inward flux transport. With more and more magnetic flux accumulating in the inner region of the disk, the MAD state can be achieved as expected. Here we briefly summarize the model calculations as follows. We adopt the cylindrical coordinate $(R,\varphi,z)$ and consider a steady ADAF around a Schwarzschild black hole. The gas within an ADAF is very hot and has a positive Bernoulli parameter, thus one may expect that a strong outflow associated with its accretion process\citep[][]{1994ApJ...428L..13N,1995ApJ...452..710N,1999MNRAS.303L...1B,1999MNRAS.310.1002S,2015ApJ...804..101Y}. Due to the high ionization state of the gas, these outflows are difficult to detect. In this work, following \cite{1999MNRAS.303L...1B}, we assume that the mass accretion rate $\dot{M}_{\rm acc}$ to be a power law dependence of the radius, which reads

\begin{equation}\label{eq:Macc}
\dot{M}_{\rm acc}(R)=\dot{M}_{\rm out}\left(\frac{R}{R_{\text {out }}}\right)^s.
\end{equation}
Here, $\dot{M}_{\rm acc}=-2 \pi R \Sigma v_{R}$ and $\dot{M}_{\rm out}$ are the mass accretion rate at radius $R$ and the outer boundary $R_{\rm out}$, respectively. The mass-loss rate within the outflow is determined by the power index $s$ which, indeed, should be a function of radius.  \cite{2009MNRAS.400.1734L} constructed a steady ADAF model and calculated the global dynamics of the accretion disk with magnetic-driven outflows. They found that the mass accretion rate $\dot{M}_{\rm acc}$, very close to a power-law $R$-dependence, only in the inner region close to the black hole $\dot{M}_{\rm acc}$ has a slight deviation. In this work we mainly focus on the magnetic flux transport and formation of a MAD state within an ADAF, the detailed nature of the outflow driven from the surface of the ADAF is out of the scope of current work. To avoid the complicated details of the outflow nature, we assume that the power index  $s$ is a constant as that adopted in \cite{2019ApJ...887..167X}. We adopted the pseudo-Newtonian potential (Paczy\'nisk-Wiita potential \cite{1980A&A....88...23P}) to mimic the gravity potential geometry of a Schwarzschild black hole, thus the Keplerian angular velocity is given by
\begin{equation}\label{eq:omegak}
\Omega_{\mathrm{K}}^2(R)=\frac{G M_{\rm BH}}{\left(R-R_{\mathrm{s}}\right)^2 R},
\end{equation}
where $R_{\mathrm{s}}=2GM_{\rm BH}/c^2$ and $M_{\rm BH}$ are the Schwarzschild radius and black mass, respectively.

As discussed above, we employ a power-law $R$-dependence to describe the mass accretion rate, thus the continuity equation reads
\begin{equation}\label{eq:continuity}
\frac{d}{d R}\left(2 \pi R \Sigma v_{{R}}\right)+\dot{M}_{\rm out}\frac{d}{d R}\left(\frac{R}{R_{\text {out }}}\right)^s=0.
\end{equation}
Here, $v_R$ and $\Sigma=2\rho H$ are the radial accretion velocity and the surface density of the accretion flow, respectively. We assume a stationary state along the vertical direction of the accretion flow, thus the half thickness of the disk is given by $H=c_{\rm s}/\Omega_{\rm k}$ with $\Omega_{\rm k}$ is the Keplerian angular velocity. $c_{\rm s}$ is the isothermal sound speed given by $c_{\rm s}^2=P_{\rm tot}/\rho$. The total pressure is contributed by the magnetic pressure $P_{\rm m}$ and the gas pressure $P_{\rm g}$, i.e., $P_{\rm tot}=P_{\rm m}+P_{\rm g}$. $\beta=P_{\rm g}/P_{\rm m}$ is the ratio of gas to magnetic pressure.

The angular momentum equation for a steady accretion flow
is
\begin{equation}\label{eq:angular}
-\frac{d}{dR}\left[\dot{M}_{\rm acc}(R^2 \Omega-j_{\rm in}) \right]=
\frac{d}{d R}\left(2 \pi R \nu \Sigma R^2 \frac{d \Omega}{d R}\right)+2\pi R T_{\rm  m} -  R^2 \Omega \frac{d \dot{M}_{\rm acc}}{d R},
\end{equation}
where $\nu =\alpha c_{\rm s}H$ is the $\alpha$ viscosity and $\Omega$ is the angular velocity of the gas within the accretion flow. $j_{\rm in}$ represent the specific angular momentum of the gas swallowed by the black hole under the zero-torque condition\citep[][]{1998tbha.conf..148N}, with the boundary conditions given at $R_{\rm out}$, the precise values of $j_{\rm in}$ are determined numerically to have a smooth transonic solution. Some simulation work suggests that in the MAD system, MRI is likely to be suppressed, the magnetic interchange instability (MII) should be responsible for the transport of angular momentum \citep[][]{2011MNRAS.418L..79T,2012MNRAS.423.3083M,2018MNRAS.478.1837M,2019ApJ...874..168W}. The work done by \cite{2022MNRAS.511.2040B}, recently, argues that MRI is not fully suppressed, they found that in the MAD state, the toroidal component dominates the poloidal component, MRI is probably responsible for the existence of a strong toroidal field, and plays an important role in the transport of angular momentum, the product of the traditional $\alpha$ parameter and the plasma $\beta$ parameter in the MAD region is almost a constant (i.e., $\alpha \times \beta=\rm const$). Despite the complex nature of the MRI and MII in the MAD region, both of them are related to the turbulent magnetic field within the accretion flow. In order to avoid taking these highly nonlinear turbulent processes into consideration, we use the traditional $\alpha$ parameter to account for the angular momentum transport in equation \eqref{eq:angular}. The last term on the right-hand side of equation \eqref{eq:angular} accounts for the angular momentum loss into the outflow. The physics of the transition region on which the gas is transferred from the inward accretion flow to the outflow is relatively complex. The angular velocity of the outflowing gas may deviate from local $\Omega$. In this work, to avoid the complexity process of the transition region, we assume that at a given radius $R$ the outflowing gas has an angular velocity the same as it in the accretion flow, i.e., $\Omega(R)$. 

The torque exerted by the global magnetic field on the unit area of the disk surface is
\begin{equation}\label{eq:Tm}
    T_{\rm m}=\frac{B_z B_{\varphi}^{\rm s}}{2\pi}R,
\end{equation}
where $B_{\varphi}^{\rm s}$ is the toroidal component of the global magnetic field at the disk surface.

Along the radial direction, the outward magnetic force caused by the large-scale magnetic field decelerates the accretion flow, in this case, the radial momentum equation reads
\begin{equation}\label{eq:angular_radial} 
v_{R} \frac{d v_{R}}{dR}-R\left(\Omega^2-\Omega_{\mathrm{K}}^2\right)=-\frac{1}{\rho} \frac{d P_{\rm tot}}{d R}+g_{\mathrm{m}},
\end{equation}
where $g_{\rm m}$ is the radial magnetic force caused by the curvature of the magnetic field line which is
\begin{equation}
g_{\mathrm{m}}=\frac{B_{R}^{\mathrm{s}} B_{z}}{2 \pi \Sigma}.
\end{equation}
Here, $B_{R}^{\mathrm{s}}$ and $B_{z}$ are the radial and vertical components of the large-scale poloidal magnetic field on the disk surface, in the derivation of $g_{\rm m}$ we adopt the approximation $\partial B_R/ \partial z \simeq B_R^s/H$  \citep[see ][for details]{2009MNRAS.400.1734L,2011ApJ...737...94C}.

The energy equation is
\begin{equation}\label{eq:energy} 
    \Sigma v_{R}\left(\frac{2c_{\rm s}}{\gamma-1} \frac{d c_{\mathrm{s}}}{d R}-\frac{c_{\mathrm{s}}^2}{\rho} \frac{d \rho}{d R}\right)=f_{\rm adv}Q_{\rm vis}^{+},
\end{equation}
where $\gamma$ is the ratio of specific heats, and $f_{\rm adv}$ is an energy advection factor, which describes how much of the dissipated energy in the disk is advected in the accretion flow. $Q_{\rm vis}^{+}$ is the viscous heating rate within the accretion flow, defined as 
\begin{equation}\label{eq:dissipation}
   Q_{\rm vis}^{+} =\nu \Sigma\left(R \frac{d \Omega}{d R}\right)^2.
\end{equation}
Basically, in the energy equation, one should consider the energy loss by radiation and that carried away by the outflow. Due to the radiation inefficient nature of ADAF, a minuscule fraction of the dissipated energy is radiated out. The energy carried away by the outflow includes two parts. First, in order to escape from the accreting system, a certain fraction of the dissipated energy should be delivered into the outflow to overcome its local binding energy. Another is the local kinetic energy of the outflow material\citep[see][for details]{1999MNRAS.309..409K}. Here we use a parameter $f_{\rm adv}$ to account for the energy loss caused by both radiation and carried away with the outflowing gas. Typically, the observation requires $f_{\rm adv}\sim 0.5-0.9$ \citep[][]{2012MNRAS.427.1580X}. The precise value of $f_{\rm adv}$, in principle, is governed by the heating (e.g., viscous heating, magnetic reconnection heating, etc.) and cooling processes (radiation, outflow, etc.).  Here, for simplicity, we adopt $f_{\rm adv}$ as a constant \citep[see also][]{2011ApJ...737...94C, 2019ApJ...887..167X}.

The large-scale magnetic field advected in an ADAF was studied in \cite{2011ApJ...737...94C}. Here we briefly summarize the model calculations as follows. Based on the above disk structures, the quasi-static evolution of the large-scale magnetic field within an ADAF can be achieved by solving the induction equation at different steady states.  Under the axisymmetric assumption, in cylindrical coordinates, the steady-state induction equation can be expressed as (see equation (10) in \cite{1994MNRAS.267..235L})
\begin{equation}\label{eq:induc_psi}
-v_{R}(R)
\frac{\partial }{\partial R}
\left[ R \psi(R,z) \right] - \frac{4 \pi \eta}{c}R J_ \phi(R,z)=0,
\end{equation}
where $\eta$ is the magnetic diffusivity which is related to the magnetic Prandtl number by ${\cal P}_{\rm m} = \nu/\eta$. The typical value of ${\cal P}_{\rm m}$  is found to be around ${\cal P}_{\rm m}\sim 1-5$ \citep[][]{2009A&A...507...19F,2009ApJ...697.1901G,2009A&A...504..309L}. In case of efficient flux transport ${\cal P}_{\rm m}\ga R/H$ is often needed, for a typical scale height of the accretion flow (e.g., $H/R\sim 0.2-0.5$), we adopt ${\cal P}_{\rm m}=3$ in all the calculations. The stream function $R\psi(R,z)$ is proportional to the magnetic flux within radius $R$ located at vertical position $z$. In the disk, the current density $J_ \varphi (R,z)$ is related to the stream function $R\psi(R,z)$ by 
\begin{equation}\label{eq:psi_Rz}
    \psi_{\rm d} (R,z)=\psi (R,z) - \psi _\infty
    =\frac{1}{c} \int_{R_{\rm in}}^{R_{\rm out}} \int_{0}^{2\pi} \int_{-z_{\rm h}}^{z_{\rm h}} \frac{J_\varphi (R',z') \cos \varphi' d\varphi' R' d R' dz'}{\left[ R^2 + R'^{2} + \left( z-z'\right)^{2} - 2R R' \cos \varphi'\right]^{\frac{1}{2}}} ,
\end{equation}
where $\psi_{\rm d} (R,z)$ is contributed by the current inside the accretion flow, $c$ is the light speed.  The strength and configuration of the large-scale poloidal magnetic field and $\psi (R,z)$ are related by

\begin{equation}\label{eq:br_bz}
\begin{aligned}
B_{R}(R,z) =& -\frac{\partial}{\partial z} \psi(R,z),\\
B_z(R,z) =& \frac{1}{R} \frac{\partial}{\partial R}\left[ R \psi(R,z) \right].
\end{aligned}
\end{equation}

Under the flux advection picture, a homogeneous external magnetic field along the vertical direction is often assumed, i.e., $B_{\rm ext}$ to be advected inward by the accretion flow, and thus we have $\psi _\infty =\frac{1}{2}B_{\rm ext}R$.
Differentiating equation \eqref{eq:psi_Rz} and substituting it into equation \eqref{eq:induc_psi} we have
\begin{equation}\label{eq:psi_Rz_diff}
-\frac{1}{c} \int_{R_{\rm in}}^{R_{\rm out}} \int_{0}^{2\pi} \int_{-z_{\rm h}}^{z_{\rm h}} \frac{\left[R'^2 + (z-z')^2 - RR' \cos \varphi' \right]J_\varphi (R',z') \cos \varphi' d\varphi' R' d R' dz'}{\left[ R^2 + R'^{2} + \left( z-z'\right)^{2} - 2R R' \cos \varphi'\right]^{\frac{3}{2}}} - \frac{4 \pi \eta}{c} \frac{R}{v_{R}(R)}  J_ \phi(R,z) = B_{\rm ext}R,
\end{equation}
in above equation, along vertical direction, we take a homogeneous distribution of the radial accretion velocity, i.e., we assume $v_{R}(R,z)=v_{R}(R,0)=v_{R}(R)$, for the detailed distribution of $v_{R}(R,z)$, one need to consider the vertical momentum equation and the radial momentum equation along vertical direction, this will enroll more complicated physics, which out of the scope of current work.  Equation \eqref{eq:psi_Rz_diff} can be rewritten as a set of linear equations,
\begin{equation}\label{eq:linear}
    -\sum_{ j=1}^{\rm n} \sum_{ l=1}^{\rm k} J_\varphi (R_{ j},z_{l}) P_{i,j,k,l} \Delta R_j \Delta z_l
    -\frac{4\pi \eta}{c} \frac{R_i}{v_R(R_i)} J_ \varphi (R_i,z_k) = B_{{\rm ext}} R_i,
\end{equation}
where 
\begin{equation}
P_{i,j,k,l}= {\frac{1}{c}}\int_{0}^{2\pi}{\frac {\left[ R_j^2 + \left( z_k - z_l \right)^2 - R_i R_j {\cos}\varphi'\right]R_j}{\left[ {R_i}^2+{R_j}^2 + \left( z_k - z_l \right)^2 -2R_i{R_j}\cos{\varphi '}\right] ^{\frac{3}{2}}}}\cos \varphi'{d}\varphi ',
\end{equation}
the subscripts $``_{i,j,k,l}"$ are labeled for the values of the parameters at radius $R_i,R_j$ and vertical position $z_k,z_l$. $J_\varphi(R_j,z_l)$ is the current density at position $(R_j, z_l)$.  Solving linear equations \eqref{eq:linear} with the given structures of the accretion flow, i.e., the radial velocity profile $v_R(R)$, the disk thickness 
 and the current density distribution within the accretion flow can be solved, and then the spatial distribution of the magnetic stream function $\psi (R,z)$. The poloidal magnetic field configurations can be derived by equation \eqref{eq:br_bz}. At surface (i.e., $z=\pm H$) of the accretion flow, the toroidal component $B_{\varphi}^{\rm s}$ of the global magnetic field related to the rotation of the accretion flow, from the usual MHD equations, one can have $B^{\rm s}_R\simeq(2\pi/c)J_{\varphi}^{\rm s}$, and $B^{\rm s}_\varphi \simeq(2\pi/c)J_{R}^{\rm s}$ (see equation (16) in \cite{1994MNRAS.267..235L} for the details), $J_{\varphi}^{\rm s}$  and $J_{R}^{\rm s}$ are the equivalent surface current density, thus a rather good approximation $B^{\rm s}_\varphi/B^{\rm s}_R\simeq J_{R}^{\rm s}/ J_{\varphi}^{\rm s}\simeq v_R/v_\varphi \sim 10^{-1}-10^{-2}$ can be made under the flux advection picture. One may expect a global field configuration not far from the surface of the accretion flow, in which the vertical component is dominant, especially in the MAD state. Note that in equation \eqref{eq:Tm} $T_{\rm m}<0$, thus we adopt $B^{\rm s}_\varphi= - 0.1 B^{\rm s}_R$ in all the calculations, see also that the rough estimation made in \cite{2019ApJ...887..167X}, $ B^{\rm s}_R$ can be solved by equation \eqref{eq:br_bz}, i.e., $B^{\rm s}_R=B_{R}(R,z)|_{z=\pm H}$ when the structures of the accretion flow are specified.

The radial magnetic force gradually enhanced, as the magnetic flux continued to accumulate, this led to the accretion flow rotating slowly down and decelerating radially, when the outward magnetic force compared to the local effective gravity, the accretion flow goes into a MAD state \citep[see][for details]{2003PASJ...55L..69N}. The empirical properties of MAD are partly established from numerical simulations, the detailed physical process leads to the magnetic flux saturation and internal force equilibrium to form a steady/quasi-steady MAD structure, however, still remains unclear. The work done by \cite{2003PASJ...55L..69N} suggested that, considering the force exerted on the accreting gas, when the outward magnetic force is strong enough to compete against the gravity exerted by the central black hole, i.e., $\frac{B_{R}^{\mathrm{s}} B_{z}}{2 \pi \Sigma} \sim \frac{GM_{\rm BH}}{R^2}$, a MAD state is achieved.  Generally speaking, at the MAD state, the inward and outward force equilibrium include another two parts, i.e., the inward effective gravity $R\left(\Omega^2-\Omega_{\mathrm{K}}^2\right)$ is counterbalanced by the outward pressure gradient force $-\frac{1}{\rho} \frac{d P_{\rm tot}}{d R}$ and the radial magnetic curvature force $\frac{B_{R}^{\mathrm{s}} B_{z}}{2 \pi \Sigma}$, thus under the simplified axisymmetric analysis,  \cite{2003PASJ...55L..69N} suggested that radial velocity of the MAD region is around $v_R\sim 0.01-0.001 v_{\rm k}$ ($v_{\rm k} $ is keplerian velocity with respect to the central black hole). But this is just a rough estimation and does not take the transonic nature of the accretion flow near the event horizon into consideration, for the MAD region the radial velocity near the horizon should be transonic with $v_R\sim 0.1-1v_{\rm k}$. And also, in the radial momentum equation \eqref{eq:angular_radial}, as the forces gradually reached a balanced state and the accretion flow goes into MAD state, the radial differential of the radial accretion velocity should be asymptotically equal to zero, i.e., $\frac{d}{dR}v_R=0$. However, this is a strictly mathematical critical condition and contradictory to the transonic nature of the accretion flow under our present axisymmetric MAD case. But we note here that, for the case of nonaxisymmetrical MAD, i.e., the MAD region is just like scattered strong ``magnetized voids" or sometimes called ``magnetic buoyant bubbles", \citep[][]{2008ApJ...677..317I,2012MNRAS.423.3083M,2019ApJ...874..168W}, the term $\frac{d}{dR}v_R \sim 0$ even $v_R \sim  0$ in the "magnetized voids" are possible,  but this is not the case of our work. The numerical simulation of the MAD and/or strongly magnetized accretion shows that the time or density weighted radial velocity is transonic around the inner marginally stable orbit, roughly around $(2-3)R_{\rm s}$ \citep[][]{2012MNRAS.423.3083M,2012MNRAS.426.3241N,2023ApJ...944..182D}, the theoretical calculation done by \citep[][]{2019ApJ...887..167X} show that the radial velocity of the nonaxisymmetrical MAD region is transonic around $5R_{\rm s}$ outside the marginally stable orbit. Therefore, considering the transonic nature and the simplified axisymmetric analysis done by \cite{2003PASJ...55L..69N}, i.e., $v_R\sim 0.01-0.001 v_{\rm k}$, we adopted a relatively less stringent MAD criteria,
\begin{equation}\label{eq:criteria}
\frac{d }{d R}v_R \lesssim 0.01 \frac{d }{d R}v_{\rm k}.
\end{equation}

\section{numerical setup}\label{sec:setup}
As discussed in section \ref{sec:model}, our model calculations including two parts:1. The structure and dynamics of the accretion flow; 2. the strength and configurations of the large-scale magnetic field. The calculations of any part require the results from another part, i.e., this is an iterative solution process, which makes the numerical calculations quite complicated. The numerical method adopted in this work is briefly described below.

1. The structure and dynamics of the accretion flow without magnetic field can be derived by solving the differential equation systems with suitable boundary conditions at $R=R_{\rm out}$, i.e., equations \eqref{eq:continuity},\eqref{eq:angular},\eqref{eq:angular_radial}and \eqref{eq:energy} are solved without considering the magnetic term $T_{\rm m}$ and $g_{\rm m}$.The initial boundary condition is adopted as described by the self-similar solution of \cite{1995ApJ...452..710N}. The numerical results must satisfy the zero-torque condition at the horizon, the transonic condition at a sonic point, and the outer boundary condition at $R_{\rm out}$. Thus we get the transonic solution of the normal (i.e., without magnetic field) ADAF, where the values of four variables $v_{R}$, $\rho$, $c_{\rm s}$, $\Omega$ are specified. 

2. Based on the disk structure derived above, the strength and configurations of the large-scale poloidal magnetic field are described by an integral-differential equation \citep[see][for details]{1994MNRAS.267..235L}, which can be linearized into a set of algebraic equations and then solved numerically, i.e., equation \eqref{eq:linear}. The outer boundary is adopted as $R_{\rm out}=1000R_{\rm s}$, and the surface of the accretion flow is defined as $z=\pm H$. In the calculation of the magnetic field, we adopt $n$ = 200 and $k$ = 40, which can achieve a good performance in accuracy. We adopt $n \times k$ grid cells distributed logarithmically along the radial direction and homogeneously along the vertical direction, respectively. The distribution of the effective current density $J_\varphi(R_i,z_k)$  within the flow can be derived by solving equation \eqref{eq:linear} with the above derived structure of the accretion flow, then the stream function $R\psi(R,z)$ in equation \eqref{eq:psi_Rz} can be specified with the derived effective current density $J_\varphi(R_i,z_k)$, derivatives of $R\psi(R,z)$ give the poloidal magnetic field configuration/strength, i.e., equation \eqref{eq:br_bz}. For the azimuthal component of the large-scale magnetic field,  as discussed in section \ref{sec:model}, we adopt $B^{\rm s}_\varphi= - 0.1 B^{\rm s}_R$ in all the calculations.

 3. Due to the presence of this large-scale magnetic field, the dynamical structure of the accretion flow changes significantly, we need to recalculate equations \eqref{eq:continuity},\eqref{eq:angular},\eqref{eq:angular_radial} and \eqref{eq:energy} and taking the magnetic term, i.e.,  $T_{\rm m}$ and $g_{\rm m}$, into consideration. This will give a new structure of the accretion flow under the influence of the large-scale magnetic field.

 The calculation started with a transonic solution of a normal ADAF, and the solution procedure needs to iterate between Step 2 and Step 3 as discussed above. The self-consistent solution of an inner MAD disk connected to an outer ADAF can be achieved when the inner disk meets the MAD criteria and the disk structure almost remains unchanged between two consecutive iterations. We find that such a solution is indeed available after several iterations of Steps 2 to 3 described above.
 
\section{results}\label{sec:results}
\subsection{the dynamics of the ADAF with an inner MAD region}
\begin{figure}
\gridline{\fig{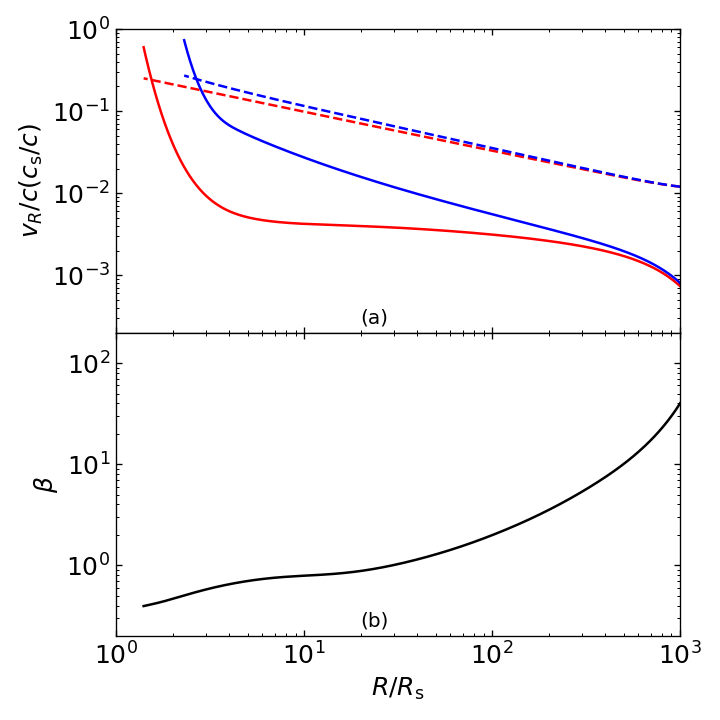}{0.5\textwidth}{}
 \fig{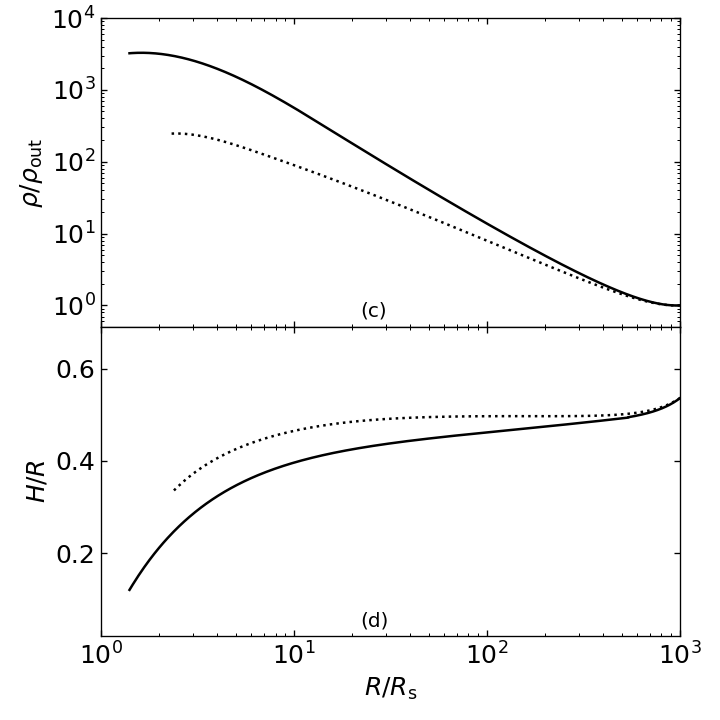}{0.5\textwidth}{}}
\caption{Dynamical structures of the accretion disk for $s=0.3$. The variations of ADAF variables with radius $R$ are shown in four panels. Panel (a): The radial
velocity (solid) and the sound speed (dashed) of the accretion flow vary with radius, the blue lines are calculated for normal ADAF without a large-scale magnetic field, and the red lines are calculated for an inner MAD disk connected with an outer ADAF. Panel (b): The radial profile of the plasma $\beta$ parameter, $\beta_{\rm out}=40$ at the outer boundary $R=R_{\rm out}$. Panel (c) and (d): The radial profile of the gas density and scale height of the accretion disk. The dotted lines are calculated for a normal ADAF, and the solid lines are calculated for the case of an inner MAD disk connected with the outer ADAF, $\rho_{\rm out}$ is the gas density at the outer boundary $R=R_{\rm out}$.
\label{fig:VrCsBeta}}
\end{figure}

\begin{figure}
    \centering
    \includegraphics[width=0.6\linewidth]{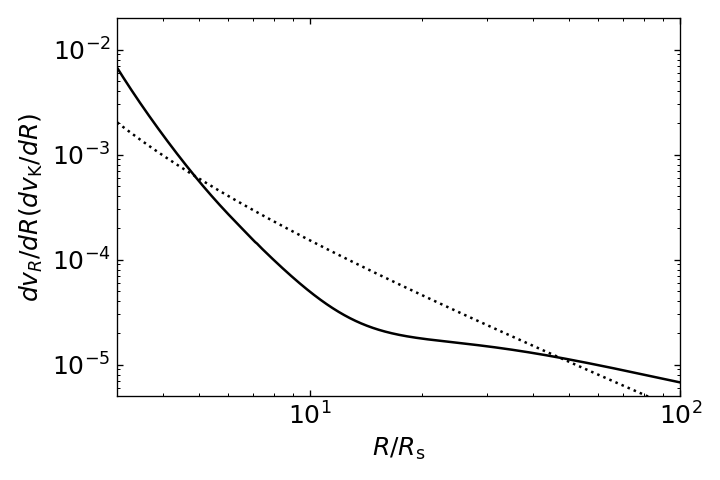}
    \caption{These lines are calculated for the criteria of the MAD state adopted in equation \eqref{eq:criteria} for the case of $s=0.3$. The solid
     line is for the derivation of the radial velocity (i.e., the solid red line in panel (a) of figure \ref{fig:VrCsBeta}) along the radial direction, while the dotted line is for 0.01 times the derivation of the Keplerian rotating velocity.}
    \label{fig:criteria}
\end{figure}

\begin{figure}
    \centering
    \includegraphics[width=0.6\linewidth]{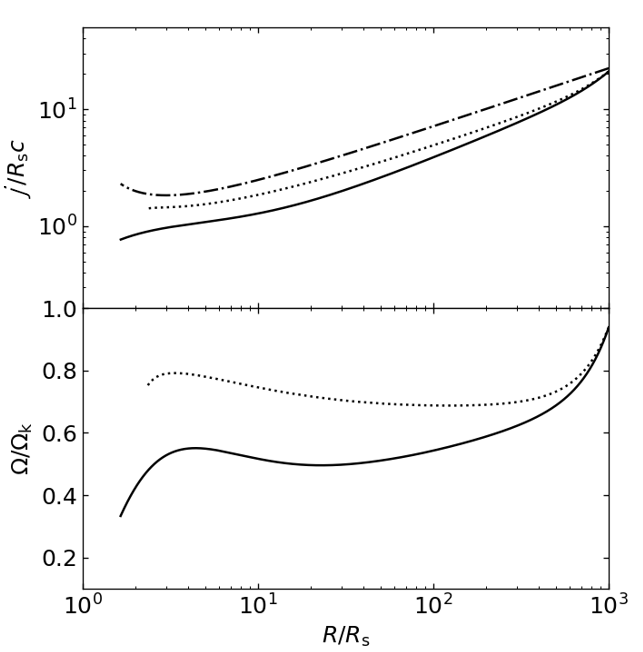}
    \caption{Radial variation of the specific angular momentum and the Keplerian rotating rate of the disk for the case of $s=0.3$. The solid and dotted lines show the variables calculated for the MAD-case and normal ADAF, respectively. The dash-dotted line is calculated for the Keplerian specific angular momentum $j_{\rm K}$}
    \label{fig:angular}
\end{figure}

\begin{figure}
    \centering
    \includegraphics[width=0.6\linewidth]{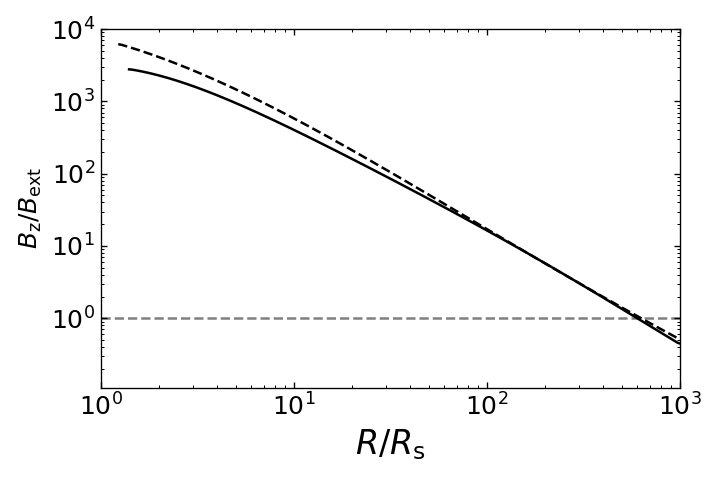}
    \caption{Variations in the vertical component of the poloidal magnetic field at disk midplane with radius. The solid line is calculated for $s=0.3$ while the dashed line for $s=0.1$}
    \label{fig:bz}
\end{figure}

we numerically solve the dynamical structure of the normal ADAF  and then derive the dynamical structure of the disk of which an inner MAD region is closely connected with an outer magnetized ADAF (hereafter referred to as MAD-case). In the calculations, we set the outer boundary of the ADAF at $R_{\rm out }=10^3 R_{\rm s}$ and the corresponding outer boundary conditions, i.e., $\rho_{\rm out}, v_{R,{\rm out}},c_{\rm s,out},\Omega_{\rm out}$, are pre-set as described by the self-similar solution in \cite{1995ApJ...452..710N}. We adopt $\alpha=0.1$ and, considering the flux advection picture to form an inner MAD region, a moderately strong external magnetic field with $\beta_{\rm out}=40$ at $R=R_{\rm out}$ is adopted in all the calculations.

\begin{figure}
\gridline{\fig{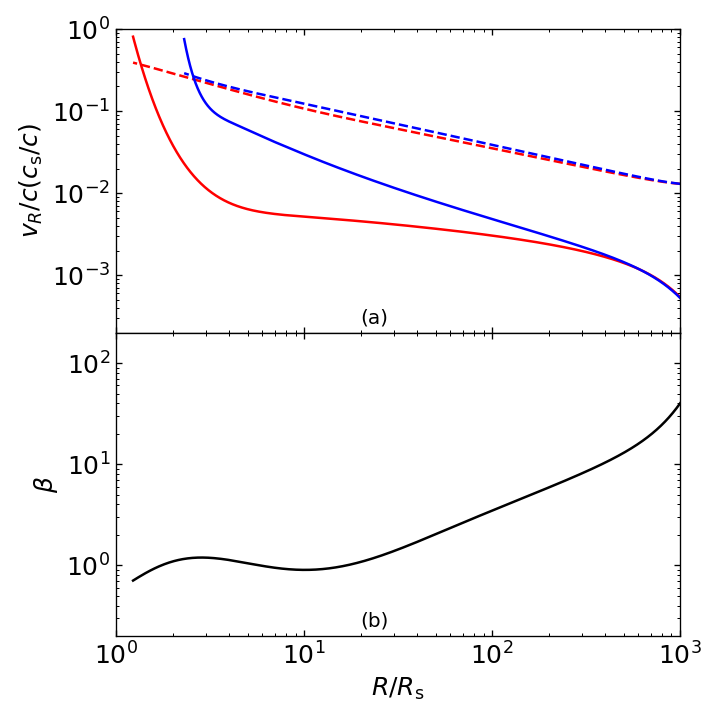}{0.5\textwidth}{}
 \fig{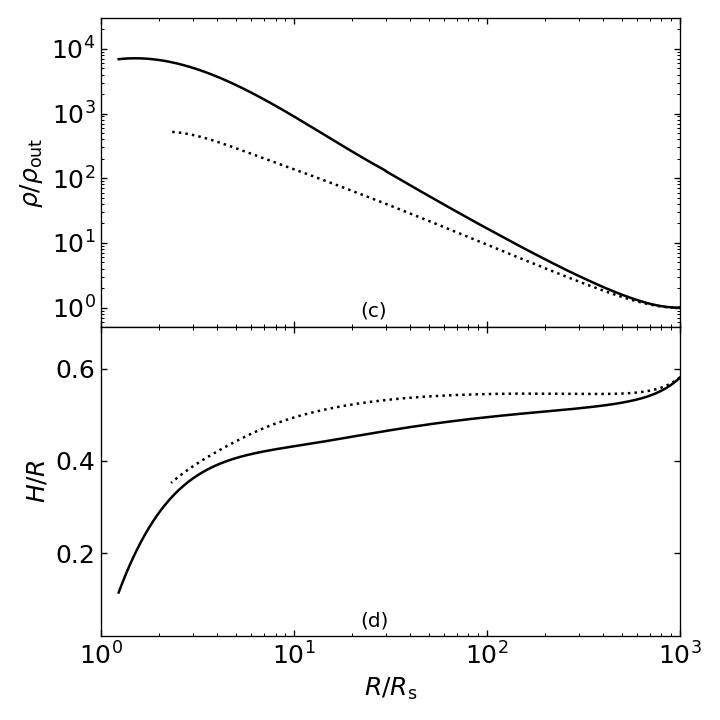}{0.5\textwidth}{}}
\caption{The same as figure \ref{fig:VrCsBeta} but the power-law index $s=0.1$ is adopted in the calculations.
\label{fig:VrCsBeta01}}
\end{figure}

\begin{figure}
    \centering
    \includegraphics[width=0.6\linewidth]{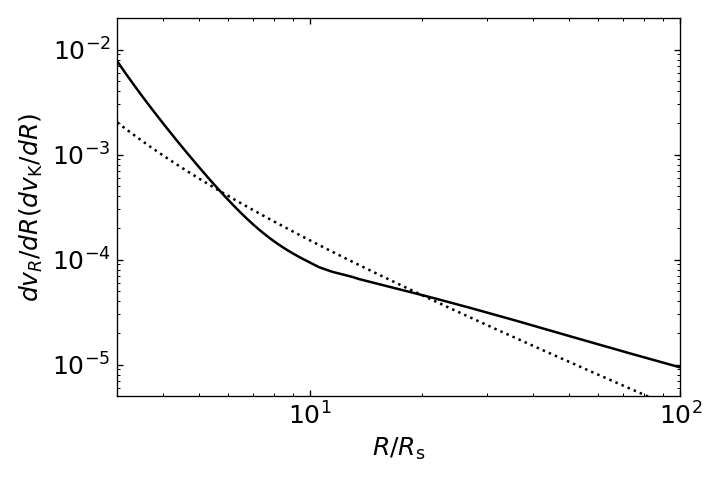}
    \caption{The same as figure \ref{fig:criteria01} but the power-law index $s=0.1$ is adopted in the calculations.}
    \label{fig:criteria01}
\end{figure}

\begin{figure}
    \centering
    \includegraphics[width=0.6\linewidth]{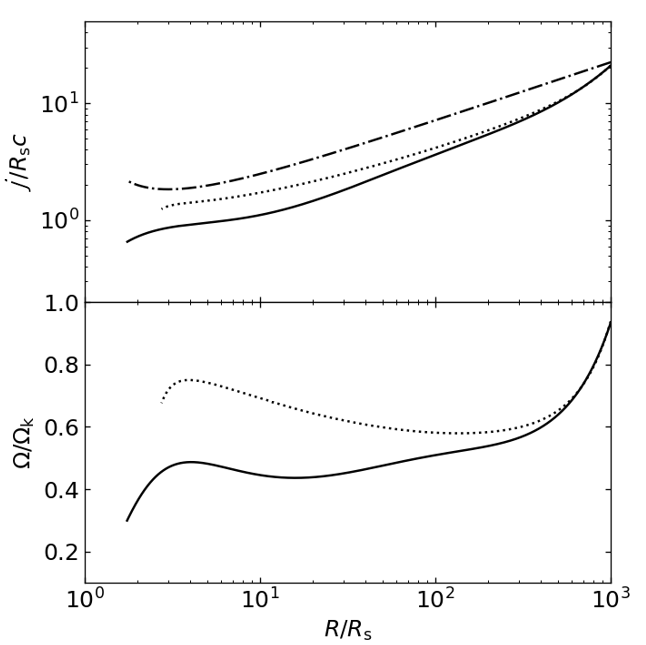}
    \caption{The same as figure \ref{fig:angular01} but the power-law index $s=0.1$ is adopted in the calculations..}
    \label{fig:angular01}
\end{figure}

In figure \ref{fig:VrCsBeta}, the dynamical structures for the accretion disk have been illustrated by adopting a constant mass loss power-law index $s=0.3$. These figures show the radial profile of radial velocity, sound speed, plasma $\beta$ parameter, disk density, and scale height, respectively. The accreting gas passes smoothly from the sonic point and falls into the black
hole. The criteria (see equation \ref{eq:criteria}) of the MAD region for the case of $s=0.3$ are plotted in figure \ref{fig:criteria} and the MAD region is about $5R_{\rm s}-45R_{\rm s}$. In the MAD region, the radial velocity of the disk is approximately 10\% that of the normal ADAF (see the solid red and blue lines in panel (a) of figure \ref{fig:VrCsBeta}). Also, we can see that the sonic point of the normal ADAF is located almost close to the marginally stable orbit $R_{\rm ms}=3R_{\rm s}$, but for the MAD-case the sonic point is located more close to the black hole horizon compared to the normal ADAF. This may result from the presence of a strong large-scale magnetic field which decreases the accretion flow radially, hence a smaller transonic radius. The MAD region is highly magnetized with $\beta \lesssim 1.0$ (see panel (b) of figure \ref{fig:VrCsBeta}). Compared with normal ADAF, in the MAD region, the gas density of the MAD-case (see the solid line in panel (c) of figure \ref{fig:VrCsBeta}) is about one order of magnitude larger than that of normal ADAF, and the corresponding scale height of the MAD region is slightly thinner than that of normal ADAF (see panel (d) of figure \ref{fig:VrCsBeta}). Due to the fact that the radial velocity of the MAD-case is significantly lower than that of normal ADAF, the slope of the gas density becomes steeper than that of normal ADAF. The upper panel of figure \ref{fig:angular} shows the radial variation of the specific angular momentum of the accretion disk. The distribution of the Keplerian angular momentum $j_{\rm K}$ is shown by the dot-dashed line for comparison. The lower panel of figure \ref{fig:angular} shows the ratio of the disk angular velocity to the Keplerian rotating rate. In the MAD region, due to the presence of the strong poloidal magnetic field, the rotating rate of the disk is highly sub-Keplerian about $0.4\Omega_{\rm K}-0.5\Omega_{\rm K}$ as well as decreasing the average specific angular momentum, $j=\Omega R^2$. 
Note that the viscosity dissipation rate, i.e., equation \eqref{eq:dissipation} depends on the radial derivation of the angular velocity. In MAD-case, due to the decrease of the angular velocity, the viscous heating within the disk is receded, hence the scale height is a little thinner compared to that of the normal  ADAF (see panel (d) in figure \ref{fig:VrCsBeta}).
From the above results, the MAD-case has a thinner (i.e., smaller $H/R$), more magnetized ($\beta\lesssim 1.0$), and more sub-Keplerian (lower $\Omega$ and smaller $j$ ) inner disk compared to the normal ADAF. The vertical component of the poloidal magnetic field is plotted in figure \ref{fig:bz}, the magnetic flux is efficiently transported inward within the ADAF, the slope of $B_z$ becomes more flatten as the radius decreases, which is caused by the decrease of the radial velocity in the MAD region compared to that of the normal ADAF.  In figures \ref{fig:VrCsBeta01}, \ref{fig:criteria01} and \ref{fig:angular01} we show the global dynamics of the normal ADAF and the MAD-case for the power-law index $s=0.1$ as a comparison. For the case of $s=0.1$, the inner MAD region is about $6R_{\rm s}-20R_{\rm s}$ (see figure \ref{fig:criteria01}), the MAD region is slightly reduced compared to the case of $s=0.3$. As the power-law index $s$ decreases, the mass accretion rate onto the black hole
increases, hence the gas density within the disk increases (see panel (c) in figures \ref{fig:VrCsBeta} and \ref{fig:VrCsBeta01}).

\begin{figure}
    \centering
    \includegraphics[width=0.6\linewidth]{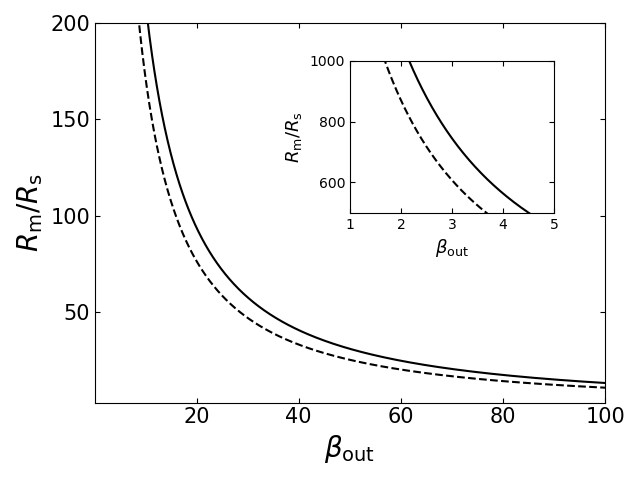}
    \caption{Variations of the inner MAD radius $R_{\rm m}$ with $\beta_{\rm out}$, in our calculations $\xi\sim -(1.45-1.55), \epsilon\sim -(1.3-1.5) $, $\tau \sim -0.42$, and $\Omega \sim (0.4-0.5)\Omega_{\rm K}$. Thus in above estimation we adopt $\xi = 1.5, \epsilon= -1.4 $, $\tau = -0.42$, $\Omega \sim 0.45\Omega_{\rm K}$, and the disk aspect ratio $\tilde{H}=0.5$ are adopted, in the MAD region the upper limit of $\beta_{\rm m}$=1.0 (solid line) and $\beta_{\rm m}$=0.5 (dashed line) are adopted in the calculations.}
    \label{fig:relation}
\end{figure}

Under the inward magnetic flux advection picture,  we can make a rough estimation on the relation of the size of the inner MAD region and the outer magnetic field strength $\beta_{\rm out}$, i.e., the ratio of the gas to magnetic pressure defined at the outer boundary $R_{\rm out}=1000R_{\rm s}$. Based on the above calculations, we can have following approximations, i.e., $B_{z}\propto R^\xi$, $\rho \propto R^\epsilon$ and $c_{\rm s}\propto R^\tau$, thus substituting the MAD criteria, (i.e., equation \ref{eq:criteria}) into equation \eqref{eq:angular_radial}, we can finally have a relation of,
\begin{equation}\label{eq:rela}
    \beta_{\rm out} \lesssim  \left[ \left(1 - \frac{\Omega^2}{\Omega_{\rm K}^2}\right) \frac{1}{\tilde{H}^2}\right]^{-1} \left[ \frac{2}{\tilde{H}} - 2\xi(\beta_{\rm m} + 1)\right]\left( \frac{R_{\rm m}}{R_{\rm out}} \right)^{2\xi-\epsilon-2\tau}-1,
\end{equation}
here, $R_{\rm m}$ is the radius of the inner MAD region, $\tilde{H}$ is the aspect ratio of the disk, $\beta_{\rm m}$ is the ratio of the gas to magnetic pressure in the MAD region at $R_{\rm m}$. The result of the above rough estimation is plotted in figure \ref{fig:relation}. The radius $R_{\rm m}$ of the inner MAD region increase with the decrease of $\beta_{\rm out}$, as the strength of the external imposed magnetic field increases, due to the efficient flux transport within the outer ADAF, the magnetic field strength significantly amplified in the inner region around the central black hole, and then an inner MAD region gradually established and expanding outward. Thus equation \eqref{eq:rela} can be useful to give a rough estimation on the size of the inner MAD region around the central black hole in some low luminosity AGNs, provided that the gas properties at the outer boundary and/or the disk dynamics have been well confined by the observations and/or numerical simulations. From figure \ref{fig:relation} we can simply infer that if one expects the entire disk transfer into a MAD state, a strong external magnetic field with $\beta_{\rm out} \sim 1-2$ is required. And also considering the transonic nature of the accretion flow and the quantity deviates from the power-law distribution in the inner region, under the picture of inward flux transport and based on our present calculations, it is unlikely to form an inner MAD disk within an ADAF when the external magnetic field is weak enough with $\beta_{\rm out} \gtrsim 100$, note that in our current work the value of $\beta_{\rm out}$ is defined at $R_{\rm out}=1000R_{\rm s}$.  There are some clues for the relation between the formation of MAD state and the magnetic field strength at the outer region of the accretion flow i.e., $\beta_{\rm out}$ at $R=R_{\rm out}$. The simulation work on the magnetic flux transport within radiatively inefficient accretion flows \citep[][]{2023ApJ...944..182D} found that, as they vary $\beta_{\rm out}$ from 70 to 7000, although a certain amount of the magnetic flux is observed to be transported into the inner region around the black hole, but none of their simulations reach the MAD state. \cite{2024ApJ...972...18A} performed a  2D  radiation-RMHD simulations on the accretion flow around spinning black hole in AGNs, they observed MAD state for  $\beta_{\rm out}=10, 25$ and SANE state for  $\beta_{\rm out}=50, 100$ in the simulations. To reach a MAD state through the in situ flux generation by turbulent dynamo, the work done by \cite{2020MNRAS.494.3656L} set a strong external magnetic field strength with  $\beta_{\rm out}=5$.  also the semi-analytical work done by \cite{2019ApJ...887..167X} found that, for the magnetic flux evolution within ADAF, $\beta_{\rm out} \sim 4$ is a near-critical SANE state, $\beta_{\rm out} \sim 2$ the inner MAD region reaches $R_{\rm m} \sim 160 R_{\rm g}$. Above works covered relatively small simulation/solution domains in about several hundred to thousand $R_{\rm g}$. Recently, the simulation work done by \cite{2020MNRAS.492.3272R} and \cite{2024arXiv240513887C} start from the Bondi radius (R $\sim 10^{5-6}R_{\rm g}$) with very weak external magnetic field strength (e.g., $\beta_{\rm out}\sim 10^6$), they found in the inner region around the black hole, the magnetic field strength can grow quickly by quasi-flux freezing, reaching $\beta \sim 1-10$. In their simulations the value of the plasma $\beta$ parameter at $R \sim 1000R_{\rm g}$ is roughly around several tens to hundreds, which is consistent with our present estimation given by equation \eqref{eq:rela}. Furthermore, these works give an insight view on the relations of the formation of the very inner MAD region and the circumstances at the Bondi radius , it is interesting to perform some further high-resolution simulations.

\subsection{estimate of the jet power for the radio galaxies}
Radio-loud AGNs are characterized by their prominent and powerful jets, which can extend far beyond the host galaxy \citep[see][for details and references therein]{2024Natur.633..537O}. For a black hole with a mass of $M_{\rm BH}$ and a dimensionless spin of $a_*$, the maximum jet power accelerated via the Blandford–Znajek mechanism \citep[][]{1977MNRAS.179..433B} for the black hole can be estimated as \citep[][]{2010ApJ...711...50T,2015ASSL..414.....C} 
\begin{equation}\label{eq:powerMAD}
    P_{\mathrm{jet} }=\frac{\kappa c}{4 \pi r_{\rm s}^2} \phi_{\mathrm{BH}}^2 \omega_{\mathrm{H}}^2 f\left(\omega_{\mathrm{H}}\right),
\end{equation}
where $\kappa$ is a constant account for the geometry for the magnetic field line, (0.053 for a split monopole geometry and 0.044 for a parabolic geometry), we adopted $\kappa=0.044$ into the calculation. $\omega_{\rm H}$ is a dimensionless parameter defined as $ \omega_{\rm H} = \Omega_{\rm H} R_{\rm s}/c$, $\Omega_H=a_*c/(2R_{\rm H})$ is the angular velocity of the black hole, which describe the inertial frame dragging effect at the horizon, $R_{\rm H} = (1+\sqrt{1-a_*})GM_{\rm BH}/c^2$ is the radius of the (outer) horizon. $\phi_{\mathrm{BH}}$ is the magnetic flux threading the horizon, i.e.,  $ \phi_{\mathrm{BH}}= \int^{\frac{\pi}{2}}_{0} \int^{2\pi}_{0} B(R_{\rm ms})R_{\rm H}^2\sin \theta d\theta d\varphi$. The correction factor $f(\omega_{\mathrm{H}})\sim 1+0.35\omega_{\mathrm{H}}^2-0.58\omega_{\mathrm{H}}^4$ is derived by using isolated jet model \citep[][]{2010ApJ...711...50T}. 
Recent numerical simulations with self-consistent accretion and jet launching \citep[][]{2023arXiv231020040C,2023arXiv231020043C}, found that the correction factor $f(\omega_{\mathrm{H}})=1$ is more favorable, thus we adopt $f(\omega_{\mathrm{H}})=1$ in above estimation. For the case of a black hole surrounded by an ADAF, the magnetic field strength $B(R_{\rm ms})$ in the inner region of the ADAF is derived from the ADAF self-similar structure \citep[][]{1995ApJ...452..710N}, see Appendix \ref{appendix} for the details. For the MAD case $B(R_{\rm ms})$ is evaluated at the marginally stable orbit, according to our calculation, the magnetic field strength in the inner MAD region has a rough relation of $B\propto R^{-\xi}$ and the power index is roughly around $\xi\sim 1.5$, thus we adopted $\xi=1.5$ in all the calculations (as that adopted in the calculation of figure \ref{fig:relation}).  The magnetic field strength threading the central region of the galaxies can be measured by using the Faraday rotation method, the observations found that the magnetic strength in the inner region of the galaxies $R\sim (10^5-10^7)\;R_{\rm s}$ is about $B\sim (0.1-10) \;{\rm mG}$ \citep[][]{2006ApJ...645..186T,2013Natur.501..391E}, thus we assume the magnetic field strength $B\sim 1\;{\rm mG}$ at $R=10^5R_{\rm s}$ to estimate the magnetic flux $\phi_{\mathrm{BH}}$ threading the horizon.

\begin{figure}
    \centering
    \includegraphics[width=0.5\linewidth]{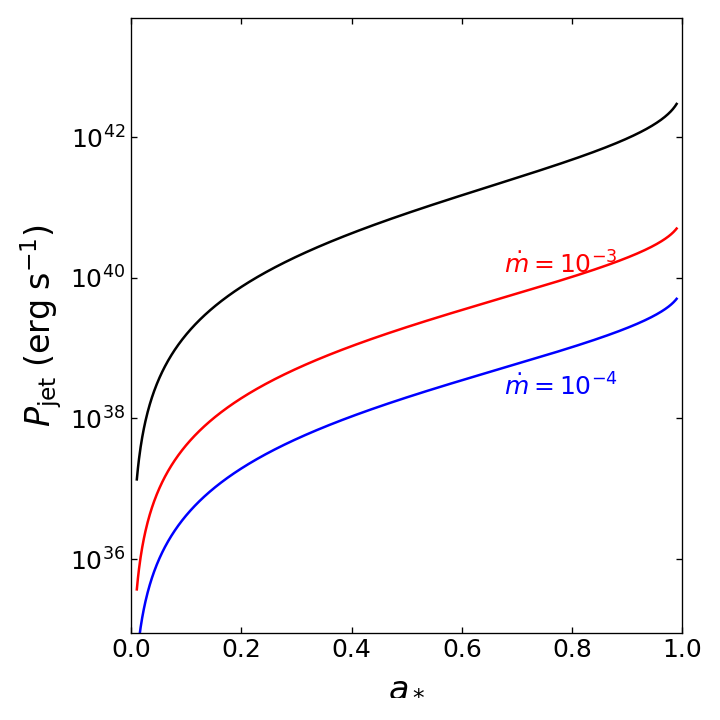}
    \caption{Variations of the jet power $P_{\rm jet}$ with the dimensionless black hole spin $a_*$ (see equation \ref{eq:powerMAD} for the details).  The dashed lines are calculated for the jet power of a black hole surrounded by a normal ADAF, while the red and blue lines corresponding to the different Eddington ratio $\dot{m}=10^{-3}$ and $\dot{m}=10^{-4}$, respectively. The solid line is calculated for the jet power of a black hole surrounded by an ADAF with an inner MAD region, the magnetic field strength of $1\;{\rm mG}$ at $R=10^5R_{\rm s}$ and $B\propto R^{-1.5}$ are used in the estimation, which we have discussed in the paragraph under equation \eqref{eq:powerMAD}. The black hole mass $M_{\rm BH}=10^8M_\odot$ is adopted.}
    \label{fig:jetspin}
\end{figure}

\begin{figure}
    \centering
    \includegraphics[width=0.5\linewidth]{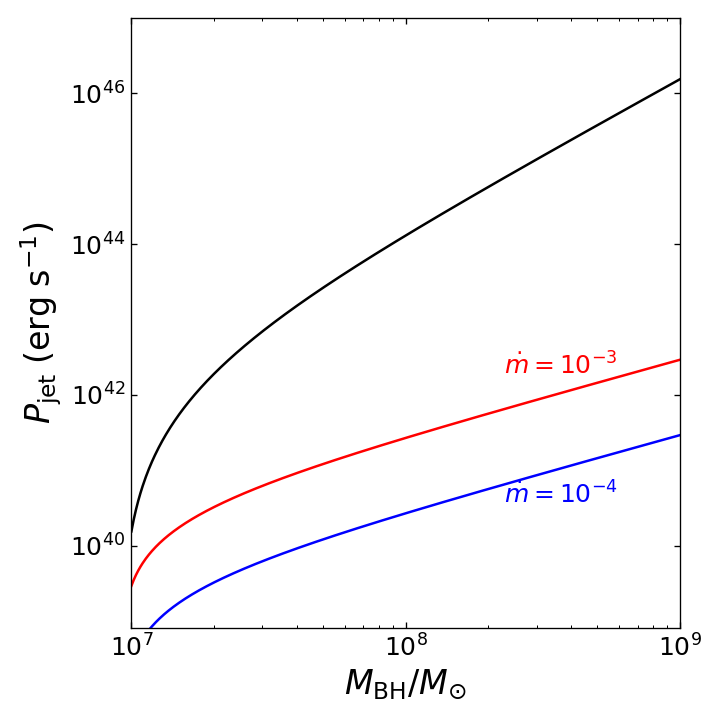}
    \caption{The same as figure \ref{fig:jetspin}, but for the variations of the jet power $P_{\rm jet}$ with black hole mass $M_{\rm BH}$. The black hole spin $a_*=0.95$ is adopted as an upper limit.}
    \label{fig:jetmass}
\end{figure}

The jet power of a black hole surrounded by an ADAF with an inner MAD region and surrounded by a normal ADAF are plotted in figure \ref{fig:jetspin}. It is obviously that for the MAD case (i.e., the solid line in figure \ref{fig:jetspin}) the jet power is about two orders of magnitude larger than that of the normal ADAF case (i.e., the dashed lines in figure \ref{fig:jetspin}).  For a fast-spinning black hole with $a_*=0.95$, we also calculate the jet power for the case of a normal ADAF and an ADAF with inner MAD for different black hole masses in figure \ref{fig:jetmass}. It is found that for a fast-spinning black hole, the jet power for the MAD case is obviously larger than that of the normal ADAF case, which is consistent with the observations of the jet power in some FR I galaxies   \citep[][]{2004MNRAS.349.1419C,2024MNRAS.530..530H}, they found that in their FR I galaxies samples with an average low Eddington ratio of $\dot{m}\sim 10^{-3.4}$, the mass accretion of their FR I sample can be modeled as ADAFs. But the estimated jet power, via the Blandford–Znajek mechanism for a fast-spinning black hole surrounded by a normal ADAF, is lower than the jet power estimated from the radio luminosity at 151 MHz for almost all the sources in their sample, This suggests a MAD around the central hole and our calculations can resolve this issue. Also for the MAD case (see the solid line in figure \ref{fig:jetmass}), the jet power increases more faster with $M_{\rm BH}$ than that of the normal ADAF case, which may caused by the reason that, due to the jet power $ P_{\mathrm{jet} }$ (see equation \ref{eq:powerMAD}) is roughly proportional to the square of the magnetic flux threading the horizon, i.e., $\phi_{\mathrm{BH}}^2$, the MAD case has a strong magnetic field strength around the horizon and as the $M_{\rm BH}$ increase the  $\phi_{\mathrm{BH}}$ increase very fast, thus the powerful jets. A large $\phi_{\mathrm{BH}}$ means the magnetic field can very efficiently extract the rotation energy of the hole to power the jets, it is 
expectably that, due to the energy extraction, the rotation of the black hole in the MAD case should be spin-down faster than that of the black hole in the normal ADAF case, this is out of the scope of our present work but worth a further study in the future.

\section{conclusions and discussions}\label{sec:cons}
The magnetically arrested disk provides an ideal dynamic environment for the launching and acceleration of powerful jets and/or fast outflows. Our aim of this work is to investigate whether a MAD disk is formed by inward magnetic flux transport within an ADAF. The magnetic flux transport within an ADAF is much more effective than that in a thin disk \citep[][]{1994MNRAS.267..235L,2011ApJ...737...94C,2019ApJ...872..149L,2023ApJ...944..182D,2024MNRAS.530.1218Y}. It can be seen that in the inner region of the ADAF, a MAD disk is formed, provided that a source of the external large-scale magnetic field exists at the outer boundary. As we know magnetic field transport within the ADAF is a global problem, the disk structure, large-scale magnetic field configuration, and the dynamics of magnetic-driven outflow are tightly coupled systems, the solution of any one of these three parts requires the other two as input conditions \citep[][]{2019ApJ...872..149L}. This work does not consider the interaction between the dynamical structure of the disk and the outflows. Similarly to many previous studies, in our work, the structure of the outflow is adopted by assuming the mass accretion rate to be a power law dependence of the radius \citep[][]{1999MNRAS.303L...1B}, therefore the calculations were simplified. Here we note that the magnetic driven outflow can significantly remove angular momentum from the disk, which helps the inward transport of the magnetic flux. So, if one takes the magnetic driven outflow into consideration, it should be easier to form a MAD disk in the inner ADAF, and also, due to sufficient angular momentum loss into the magnetic outflow, the radius of the transonic point of the accretion flow should move outward beyond the marginally stable orbit $R_{\rm ms}$. The radial velocity of the MAD region is relatively flat with radius, which means our MAD criteria equation \eqref{eq:criteria} is a pretty good approximation for the force equilibrium within the MAD region, i.e., $\frac{d}{dR}v_R\sim 0$, see the details in the paragraph above equation \eqref{eq:criteria}, at the very inner region close to the radius of marginally stable orbit, the velocity varies more steeply with radius this is caused by the fact that near the black horizon, the radial velocity of the gas is expected to have $v_R\sim v_{\rm ff}$, thus the radial velocity has to transition from the impeded MAD region to the free-fall value, see also some previous works \citep[][]{2012MNRAS.423.3083M,2012MNRAS.426.3241N,2019ApJ...887..167X,2023ApJ...944..182D}, note here that, within the marginally stable orbit, we do not consider the physics of a large number of magnetic field lines are piled together on the black horizon and the corresponding general relativistic effect, which may affect the dynamics of the accretion flow in this region, hence the radial velocity. In this work we mainly focus on the magnetic flux advected within an ADAF to form an inner MAD disk, the inward flux dragging picture is derived from the seminal model suggested by  \cite[][]{1994MNRAS.267..235L}, the outward angular momentum transport within the accretion flow is contributed by the turbulence viscosity and the outflow. We do not consider other processes which may affect the dynamics of the accretion flow, such as the magnetic buoyancy, magnetic stochasticity, magnetic reconnection, and turbulent pumping \citep[][]{2013Natur.497..466E,2018ApJ...854....2J,2018ApJ...860...52J,2019PhRvE.100a3201J}. Also, MAD accretion flows are known to show a lot of time-dependent dynamics such as magnetic flux eruptions, during which powerful disk winds are launched \citep[][]{2022ApJ...941...30C}, our recent work model is time-independent, it worth a further study on the time evolution of the magnetic flux within an ADAF and the process of inner MAD formation. \cite{2019ApJ...887..167X} set up a model to study the radiation properties of accretion flow in MAD 
state, for calculating the dynamical structures of the accretion flow, they set the global magnetic field as an input parameter. We know that the global magnetic field configuration is tightly coupled with the structure of the accretion flow, thus the global magnetic field in our work is solved based on the dynamic structure of the accretion flow, and through a series of calculations by iteration, the global field configurations and the dynamics structure of the accretion flow are self-consistent, i.e., the global field configurations can be derived using current dynamical structure of the accretion flow, at the same time under the influence of the global field, the accretion flow just stay in current dynamical structure (See \ref{sec:setup} for details).

We find that the magnetic flux can be efficiently transported within the ADAF, and eventually form a MAD disk in the inner region of the ADAF. The MAD region is about several ten $R_{\rm s}$ outside the event horizon and is highly sub-keplerian with $\Omega \sim (0.4-0.5)\Omega_{\rm K}$. The ratio of gas to magnetic pressure in the MAD region is about $\beta \lesssim 1$, which is in agreement with some previous works \citep[][]{2019ApJ...887..167X,2020MNRAS.494.3656L,2022ApJ...941...30C,2023MNRAS.521.4277R}. We make a rough estimation on the relation of inner MAD region $R_{\rm m}$ and the strength of the outer imposed magnetic field $\beta_{\rm out}$, as the external magnetic field strength increase, i.e., a lower $\beta_{\rm out}$, the region of the inner MAD disk increase, i.e., a larger $R_{\rm m}$. When the external magnetic field strength is weak to a certain extent, i.e., $\beta_{\rm out}\gtrsim 100$, MAD is unlikely to be formed through the inward flux advection process.  Due to the presence of a strong poloidal magnetic field, the radial velocity of the MAD region is about one order of magnitude smaller than that of normal ADAF. The power of the jets accelerated by a fast-spinning black surrounded by an ADAF with an inner MAD are about two orders of magnitude larger than that of the same black hole surrounded by a normal ADAF, this is in good agreement with the powerful radio jets observed in some low Eddington ratio FR I galaxies, and also the structure of an ADAF connect with an inner MAD region may be relatively common in the the center of the radio-loud AGNs. It is found that for the same outer boundary conditions, as the outflow intensity increases, i.e., the power index increases from $s= 0.1$ to $s=0.3$, the size of the inner MAD region also expending. This means that the outflow may play an important role in the formation of a MAD disk, the details of how the magnetic outflow will contribute to the formation of a MAD disk are worth further study in future work. 

\acknowledgments
The authors thank the referee for valuable comments and suggestions. This work is supported by the NSFC (12073023, 12233007, 12361131579, 12347103, and 12303020), the science research grants from the China Manned Space Project with No. CMS-CSST- 2021-A06, and the fundamental research fund for Chinese central universities (Zhejiang University), the Yunnan Fundamental Research Projects (NO.202401CF070169), the Xingdian Talent Support Plan - Youth Project, and the start-up grant from Yunnan University.


\appendix

\section{Details for the Calculation of the jet power around ADAF}
\label{appendix}
Here we list the equations and parameters that we have used in section \ref{sec:results} to estimate the jet power of the black surrounded by a normal ADAF. The structures of the self-similar ADAF are derived from \citep[][]{1995ApJ...452..710N}. Here the black hole mass scaled by the solar mass, i.e., $m=M_{\rm BH}/M_\odot$, the Eddington ratio is defined as $\dot{m}=M_{\rm acc}/M_{\rm Edd}$ with $M_{\rm Edd} $ is the Eddington accretion rate:

\begin{equation}
 B=6.55 \times 10^8 \alpha^{-1 / 2}(1-\frac{\beta}{1+\beta})^{1 / 2} c_1^{-1 / 2} c_3^{1 / 4} m^{-1 / 2} \dot{m}^{1 / 2} \tilde{R}^{-5 / 4}\;\; \mathrm{G}.
\end{equation}
Where $\tilde{R}=R/R_{\rm s}$ ($R_{\rm s}={2GM_{\rm BH}}{c^2}$), and $\beta = P_{\rm g}/P_{\rm m}$ is the ratio of gas to magnetic pressure. The constants $c_1$, $c_2$, $c_3$ are defined as 

\begin{align*}
c_1 & =\frac{5+2 \varepsilon^{\prime}}{3 \alpha^2} g^{\prime}\left(\alpha, \varepsilon^{\prime}\right), \\
c_2 & =\left[\frac{2 \varepsilon^{\prime}\left(5+2 \varepsilon^{\prime}\right)}{9 \alpha^2} g^{\prime}\left(\alpha, \varepsilon^{\prime}\right)\right]^{1 / 2}, \\
c_3 & =\frac{2\left(5+2 \varepsilon^{\prime}\right)}{9 \alpha^2} g^{\prime}\left(\alpha, \varepsilon^{\prime}\right), \\
\varepsilon^{\prime} & \equiv \frac{1}{f}\left(\frac{5 / 3-\gamma}{\gamma-1}\right), \\
g^{\prime}\left(\alpha, \varepsilon^{\prime}\right) & \equiv\left[1+\frac{18 \alpha^2}{\left(5+2 \varepsilon^{\prime}\right)^2}\right]^{1 / 2}.
\end{align*}

all the quantities used to calculated the jet power are estimated at, $R_{\rm ms}$, the marginally stable orbit of the accretion flow \citep[see][for details]{1972ApJ...178..347B}, i.e.,
\begin{align*}
R_{\mathrm{ms}} & =\frac{R_{\mathrm{s}}}{2}\left\{3+Z_2 - \left[\left(3-Z_1\right)\left(3+Z_1+2 Z_2\right)\right]^{1 / 2}\right\}, \\
Z_1 & \equiv 1+\left(1-a_*^2\right)^{1 / 3}\left[\left(1+a_*\right)^{1 / 3}+\left(1-a_*\right)^{1 / 3}\right], \\
Z_2 & \equiv\left(3 a_*^2+Z_1^2\right)^{1 / 2},
\end{align*}
note here that in order to estimate the maximum jet power, we adopt prograde orbit for the accretion flow with respect to the spin of the central black hole.


\bibliography{ms}{}
\bibliographystyle{aasjournal}



\end{document}